![frontiers]

# A two-scale FEM-BAM approach for fingerpad friction under electroadhesion

**Fabian Forsbach**[1*], **Markus Heß**[1], **Antonio Papangelo**[2,3]

[1]Institute of Mechanics, Technische Universität Berlin, Berlin, Germany

[2]Department of Mechanics Mathematics and Management, Politecnico di Bari, Bari, Italy

[3]Department of Mechanical Engineering, Hamburg University of Technology, Hamburg, Germany

**\* Correspondence:**
Fabian Forsbach
fabian.forsbach@tu-berlin.de

**Abstract**

The complex physics behind electroadhesion-based tactile displays poses an enormous modeling challenge since not only the fingerpad structure with multiple nonlinear layers, but also the roughness at the microscopic scale play a decisive role. To investigate tactile perception, a potential model should also offer the possibility to extract mechanical stimuli at the sites of the relevant mechanoreceptors. In this paper, we present a two-scale approach that involves a finite element model (FEM) at the macroscopic scale and a simple bearing area model (BAM) that accounts for the measured roughness on the papillary ridges. Both separate scales couple in an iterative way using the concept of an equivalent air gap. We show that the electroadhesion-induced changes in friction and contact area predicted by the proposed model are in qualitative agreement with recent experimental studies. In a simple example, we demonstrate that the model can readily be extended by a neural dynamics model to investigate the tactile perception of electroadhesion.

## 1  Introduction

By integrating the principle of electrovibration into a capacitive touch panel, Bau et al. (2010) introduced an innovative technology for enhancing touch interfaces with tactile feedback. This technology modulates the friction between the sliding finger and touch surface by electrostatic actuation. The latter arises from an alternating voltage applied to the conductive layer of the touchscreen, which generates an attractive force between the finger and touch surface. Controlling the amplitude, frequency and waveform of the input voltage enables rendering virtual shapes and textures on the physically flat touch surface (Kim et al., 2013; Osgouei et al., 2017; Wu et al., 2017). The great potential of this innovative, easy-to-integrate, and low-power technology has stimulated extensive research over the past decade which is still ongoing. Some distinguished works have emerged both in the experimental field and in theoretical as well as numerical modeling, which contributed significantly to a better understanding of the predominant physical processes in the frictional contact interface between the finger and touch surface. Experiments include measurements of the frictional force at different amplitudes, shapes, and frequencies of the voltage input signal, as well as psychophysical experiments investigating the tactile perception of electrovibration. From the measurements by Meyer et al. (2013), a quadratic dependency of the frictional force (and the inferred electrostatic contribution to the normal contact force) on the amplitude of the electrical voltage emerged, which confirmed the applicability of the simple parallel-plate capacitor model. Furthermore, in the prescribed frequency range from 10Hz up to 10kHz they observed an increase in the inferred electrostatic normal force with increasing frequency of a sinusoidal input signal. In a recent publication, Aliabbasi et al. (2022) extend the investigated range of stimulation frequencies from 1Hz to 1MHz and identify two prominent peaks at 250Hz and 100kHz. They conclude that the frequency-dependent behavior of the electrostatic attraction force at frequencies below the first peak of 250Hz is dominated by charge leakage, while the frequency-dependent electrical properties of the stratum corneum (SC) primarily influence the behavior above 250Hz. Shultz et al. (2015) proposed a RC impedance model with an interfacial air gap impedance as its central part to map the frequency-dependent electrostatic attraction between the finger and touchscreen. This model assumes that the only friction-relevant voltage is the one across the small interfacial gap between the surfaces of the finger and touchscreen. Due to their roughness the interfacial air gap is non-uniform, and the real contact area is typically a small fraction of the nominal contact area, since only asperities on the microscale make contact. In a later experimental study (Shultz et al., 2018), the same authors showed that the interfacial gap impedance is significantly lower for the stationary finger in comparison to a sliding one and hypothesized that the lower impedance in the stationary case was due to the accumulation of sweat in the air gap. Ayyildiz et al. (2018) have experimentally investigated the influence of both the externally applied normal force and the voltage amplitude on the frictional force between the finger and the touchscreen. In addition, they have applied Persson's mean field theory based on multiscale contact mechanics (Persson, 2018), which predicted results that were in good agreement with the measured data when parameters were appropriately fitted. A much simpler theory for electroadhesion between rough surfaces was proposed by Ciavarella and Papangelo (2020). It is based on the bearing area model (BAM) introduced by Ciavarella (2018) and later extended for rough surfaces with high fractal dimension by Ciavarella and Papangelo (2019). According to BAM, the solution of the adhesive contact is composed of a repulsive non-adhesive solution and an attractive adhesive part which can be found separately. Like in the DMT theory adhesive forces act only in a non-contact area outside of the compressive contact area and do not deform the contact shape. However, a simple Maugis-Dugdale law of attraction is assumed, and the area of attraction is estimated by the change of the bearing area geometrical intersection when the indentation is increased by the corresponding Maugis-Dugdale range of attraction. In the macroscopic modeling of Heß and Forsbach (2020), Shull's compliance method is applied which assumes adhesive interactions only within the contact area in terms of an interfacial binding energy. It requires the solution of the corresponding non-adhesive contact problem, for which the power functions given by Dzidek et al. (2017) were taken, originating from a fit to experimental data. To account for the whole influence of the non-uniform interfacial air gap the concept of an equivalent air gap introduced by Heß and Popov (2019) was applied. It goes without saying that the thickness of the equivalent air gap in principle depends on both the externally applied force and the amplitude of the applied voltage. However, Heß and Forsbach neglected the influence of voltage and assumed a pure linearly decreasing function of the external load instead. By adjusting the two polynomial coefficients of this linear approach for the equivalent air gap, they fitted the voltage-dependent frictional force that emerged from their model to agree well with the experimental data of Ayyildiz et al. (2018). A common flaw of all the above-mentioned theoretical approaches is that they assume linear elasticity (and sometimes plasticity). This may be partially justified for the study of skin deformation in the near-surface region and from a microscopic point of view, but stresses and deformations inside the finger cannot be correctly captured. The skin has a specifically layered structure and individual layers exhibit highly non-linear material behavior. Furthermore, the junctions of the layers have a functionally based special geometry. Macroscopic processes of the fingerpad, such as a preceding small rolling motion when a tangential load is applied to a finger contacting the screen, can also not be accounted for by the theories.

For such purposes, the finite element method is ideal. In particular, the ability to record mechanical stress and deformation states inside the skin, plays a key role in the process of tactile perception. The mechanical skin stimuli resulting from the frictional contact (such as pressure, stretching or vibration) initially lead to a change in the state of stress and deformation at the spatially distributed mechanoreceptors in the skin layers. The latter can be activated by this, i.e., they may convert the time-varying mechanical state into neuronal impulses (action potentials). These are transmitted via further neuronal structures to the somatosensory cortex, where they are translated into a tactile sensation. Although numerous research papers have addressed this topic (Maeno et al., 1998; Shao et al., 2010; Gerling et al., 2014), electrovibration has not yet been considered in this context. To the best of the authors' knowledge, there are only two relevant papers that have even addressed incorporating electrostatic forces into a finite element simulation for mapping finger-touchscreen contact. Papangelo et al. (2020) studied the electroadhesive contact between a conductive sphere with a rigid substrate, both coated with an insulating layer. They compare results of a DMT-approach that neglects the deformations due to adhesive tractions with the one obtained from a full iterative FE analysis that accounts for such deformations. However, they assume a priori that the effective insulating layer thickness is much greater than the RMS surface roughness so that the contribution of the surface roughness on the gap function can be neglected and the contact bodies can be considered as smooth. This condition is fulfilled for industrial applications in robotics such as soft grippers, but not for the contact between finger and touchscreen. In the work of Vodlak et al. (2016), there is essentially a debate about the "right" approach for the electrostatic attraction force based on the parallel-plate capacitor theory to model electrovibration. By comparison with their own experimental results as well as those taken from the literature, they demonstrated that the well accepted formula found in textbooks is not suitable for modeling electrovibration and that another should be preferred instead. However, they ignored the roughness on the microscale, which influences the electrostatic attraction quite decisively, since most important contributions to the electrostatic attraction result only from the very small areas of real contacts and from the so-called rim-areas around where the interfacial surface separation is very small (Ayyildiz et al., 2018). In their work, the authors also develop a two-dimensional FE model of the finger pad including ridges (but again under the mentioned flaw of ignoring the roughness on smaller length scales), and implement electrostatic attraction based on the parallel plate capacitor assumption within finite contact elements at the boundary. They demonstrated how this multi-physics model can be used to render a sensation on a flat haptic screen that is equivalent to the one perceived during sliding of the finger over a textured surface. For this purpose, they calculated the required voltage profile, i.e., function of voltage in terms of central position of the finger pad, to achieve the same friction in both cases.

In the present work, we propose a two-scale approach that addresses the modeling challenges described above. The roughness on the papillary ridges, which we characterized in topographic measurements, is treated as a separate scale. This allows the determination of the equivalent air gap thickness in terms of the external loading and the applied voltage via BAM. At the macroscopic scale, which extends to the papillary ridges, a FE model with realistic geometry and material behavior is used. The electrostatic attraction realized by surface forces in the FE model couples both scales in an iterative way.

Numerous recent publications provide data from psychophysiological experiments on the perception of virtual shapes or textures generated by electrovibration (Vardar et al., 2017; İşleyen et al., 2020). Usually, a comparison with the perception of real textures is made as well. However, to the best of the authors' knowledge, no one has yet made this comparison with regard to the stress quantities like the strain energy density (SED) at the sites of mechanoreceptors, which are responsible for their stimulation and thus initiate the process of tactile perception. Our model offers this possibility, which is demonstrated by an initial example in the discussion on tactile perception of electroadhesion in Chapter 5.

## 2 Finite element model with electroadhesion

We developed the two-dimensional finite element model (plane strain) depicted in Figure 1 using the software ABAQUS for quasistatic simulations of the fingerpad in electroadhesive contact with a touchpad. Fingerpad width and height as well as the dimensions and positions of bone and nail are the same as in Shao et al. (2010) and are summarized in Table 1. The bone, where the external normal force $f_{\text{ext}}$ and tangential force $f_T$ per unit length are applied, is much stiffer than the surrounding tissues and is thus assumed to be rigid. The bone can be moved in both, normal and tangential direction relative to the touchscreen, while the rotation is constrained. To ensure a realistic connection of nail and bone, a comparatively stiff linear elastic nail bed is introduced similar to Somer et al. (2015) using the elastic properties for nail and nail bed listed in Table 2. The touchscreen is modelled as a smooth analytical rigid surface.

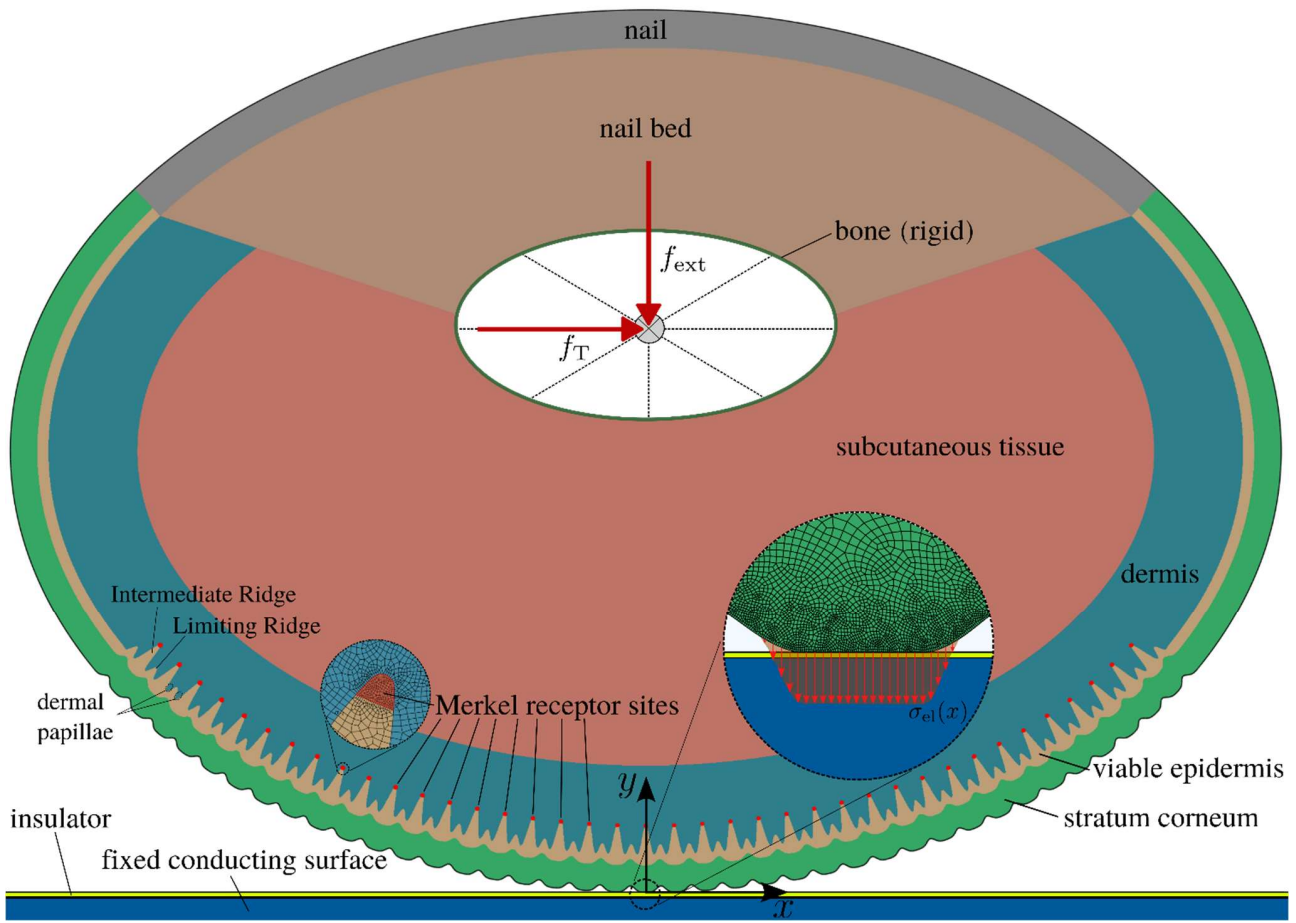

**Figure 1: Finite element model of the fingerpad touchscreen contact with the material layers. Detail views show the refined mesh at the surface with the electrostatic attraction and the refined mesh at the Merkel receptor sites.**

We used ABAQUS Explicit, due to advantages in numerical stability and efficiency over the implicit time integration scheme for this contact problem. Moderate mass scaling was employed to increase the stable time increment while maintaining the desired quasistatic character by keeping the ratio of kinetic to internal energy below 1% throughout the simulations. The electroadhesive interaction (see Section 2.2) requires a significant refinement of the mesh towards the surface. Additionally, to monitor the influence of electroadhesion on the tactile receptors, refinement at the sites of the Merkel receptors is introduced to capture local mechanical entities relevant for mechanotransduction like the strain energy density (SED). The refined mesh is shown in Figure 1. In total, ~260k four-node plane strain elements (CPE4R) are used with the smallest element dimensions of ~3µm at the contacting surface and the receptor sites.

## 2.1 Modelling of the soft tissue layers.

The soft tissues layers of skin exhibit a highly nonlinear elastic response. The layers were modelled as hyperelastic materials according to a recent study by Boyle et al. (2019) using the first order Ogden strain energy potential

$$W = \frac{2G_0}{\alpha^2} \left( \overline{\lambda}_1^{-\alpha} + \overline{\lambda}_2^{-\alpha} + \overline{\lambda}_3^{-\alpha} - 3 \right) + \frac{1}{D} (J-1)^2, \tag{1}$$

**Table 1: Geometric parameters used for the FE model**

| Parameter Name | Value and Unit |
|---|---|
| Width and height of fingertip | $W = 20$mm, $H = 14$mm |
| Thickness of stratum corneum | 425µm |
| Thickness of viable epidermis | ~175µm |
| Thickness of dermis | ~1400µm |
| Ridges (wavelength, peak to peak) | $\lambda \approx 483$µm, $A = 100$µm |
| Junction SC-VE (wavelength, peak to peak) | $\lambda \approx 467$µm, $A = 100$µm |
| Junction VE-D (wavelength, peak to peak) | $\lambda = 459$µm, $A_1 = 150$µm, $A_2 = 450$µm |

with the change in volume $J$ (determinate of the deformation gradient), the principal stretches $\overline{\lambda}_i = J^{-1/3}\lambda_i$ and the compressibility parameter $D$. Assuming incompressibility for all layers, Boyle et al. (2019) give only the initial shear modulus $G_0$ and the exponent $\alpha$. In the current model using ABAQUS Explicit, some degree of compressibility is needed for numerical stability. Thus, we assumed almost incompressible materials with $\nu = 0.48$. The compressibility parameter $D$ refers to the initial shear modulus and the Poisson's ratio by

$$D = \frac{3(1-2\nu)}{G_0(1+\nu)}. \tag{2}$$

All material parameters are summarized in Table 2.

The outermost skin layer, the stratum corneum (SC), is the stiffest of the skin layers. For plantar skin, it is 16 times thicker than for non-plantar skin (Boyle et al., 2019) and varies significantly between different subjects. For this model, we adopt the midrange value of 425µm from Jobanputra et al. (2020). The papillary ridges of plantar skin have a peak to peak amplitude of approximately 100µm. We extracted the ridge profile shown in Figure 1 by curve fitting to the measurements described in Section 3.1. The junction to the next deeper skin layer, the viable epidermis (VE), exhibits an equivalent waviness (Boyle et al., 2019). Viable epidermis is softer than stratum corneum and, for plantar skin, also significantly thinner. The junction to the next layer, the dermis (D), is more complicated and not as regular. However, it can be approximated by larger *intermediate ridges* that mirror the papillary ridges and smaller *limiting ridges* (Maeno et al., 1998; Gerling and Thomas, 2008). Many tactile receptors are located adjacent to this junction, such as the Merkel cells in tips of the intermediate ridges (see Figure 1) and the Meissner corpuscles in the dermal papillae which are located in the space in between intermediate and limiting ridges. This will be especially relevant for upcoming studies, where the influence of electroadhesion on mechanotransduction will be addressed. Note that the dermis is much softer than the viable epidermis and thus the microstructure of the junction will influence the relevant mechanical quantities at the receptor sites. The thicknesses of dermis and viable epidermis are again adopted from Jobanputra et al. (2020). All geometric details are summarized in Table 1.

**Table 2: Material parameters used for the FE model (from** Somer et al. (2015) and Boyle et al. (2019)**)**

| Layer | Material model | Values |
|---|---|---|
| Subcutaneous tissue | Ogden | $G_0 = 25\text{kPa}, \alpha = 5, \nu = 0.48$ |
| Dermis | Ogden | $G_0 = 2.55\text{kPa}, \alpha = -14.53, \nu = 0.48$ |
| Viable Epidermis | Ogden | $G_0 = 61.75\text{kPa}, \alpha = -14.53, \nu = 0.48$ |
| Stratum Corneum | Ogden | $G_0 = 86.76\text{kPa}, \alpha = -14.53, \nu = 0.48$ |
| Nail bed | linear elastic | $E = 1\text{MPa}, \nu = 0.3$ |
| Nail | linear elastic | $E = 170\text{MPa}, \nu = 0.3$ |

### 2.2 Electroadhesion: Coupling to a microscale roughness model

The electrostatic attraction between finger and touchscreen for purely capacitive behavior of the layers (that is for high frequencies of the AC input voltage $f > 1000\text{Hz}$ (Forsbach and Heß, 2021)) is given by the well-known formula

$$\sigma_{el}(x) = \frac{\varepsilon_0 U^2}{2} \left( h_0 + g(x) \right)^{-2},$$ (3)

where $\varepsilon_0$ and $U$ denote the permittivity of free space and the applied voltage, respectively, $g(x)$ is the gap function in the deformed configuration (the relative permittivity of the air gap is assumed to be $\varepsilon_{r,a} \approx 1$) and $h_0$ the thickness of the effective insulating layer $h_0 = d_{sc}/\varepsilon_{r,sc} + d_i/\varepsilon_{r,i}$. For the current model, the relative permittivity of the SC is chosen to $\varepsilon_{r,sc}(f \approx 3000\text{Hz}) = 1650$ (Yamamoto and Yamamoto, 1976), and the thickness $d_i$ and relative permittivity $\varepsilon_{r,i}$ of the insulating layer on the touchscreen are chosen according to Ayyildiz et al. (2018). The parameters of the electrostatic interaction are summarized in Table 3.

Based on Eq. (3), we developed a custom user-defined FORTRAN subroutine VDLOAD. This subroutine is called in each time increment and computes an attractive force for each node of the surface based on its current gap that is applied in the next time increment. Due to the explicit solver using many time increments and a "smooth" application of all loads, this results in the correct equilibrium state.

From previous studies (Heß and Forsbach, 2020), it is known that the microscale roughness on the papillary ridges has a strong influence on the magnitude of the electroadhesive effect. This means, setting $g(x) = 0$ for contacting nodes in a macroscopic model like the current results in unreasonably high electrostatic attraction. Thus, we adopt the concept of an *equivalent air gap* recently proposed by Heß and Popov (2019) in the following. Let us assume that there is a given function of the equivalent air gap in terms of the external normal loading on a single ridge contact. Figure 2 shows the contact pressure (blue) and the electrostatic attraction in contact (red) for different load cases. Note that the resulting elastic stresses (the difference of contact pressure and electrostatic attraction) are tensional towards the contact edges of the ridges. The external pressure $p_{ext}$ on a ridge in contact is defined as the integrated difference of the total contact pressure $p$ and the electrostatic attraction $\sigma_{el}$ in relation to the contacting length of the ridge $L_R$,

$$\bar{p}_{ext} = \frac{1}{L_R} \int_{\sigma_{el} > 0} \left[ p(x) - \sigma_{el}(x) \right] \mathrm{d}x. \tag{4}$$

The pressure in each ridge at the current time increment can be obtained using another subroutine named VFRIC that is also used to define the friction model described in the following section. The subroutines VFRIC and VDLOAD can communicate using common block arrays.

Finally, the gap for a node with current coordinates $(x_i, y_i)$ which is in vicinity of ridge $j$ with the average external pressure $\bar{p}_{ext,j}$ is determined by

$$g(x_i) = \begin{cases} d_{a,eq}\left( \bar{p}_{ext,j} \right), & y_i < d_{a,eq}\left( \bar{p}_{ext,j} \right) \\ y_i & , \text{else} \end{cases}. \tag{5}$$

The equivalent air gap that was determined with the simple microscale model described in Section 3.2 is also included in Figure 2 (green lines). As expected, the equivalent air gap is smallest for ridges in the center where the pressure is highest.

## 2.3 Friction law

Although the exact ridge geometry is used, the current FE model is still a macroscopic model that does not resolve the actual multi-asperity contact. In a recent study, the authors have shown that from a macroscopic point of view, a pressure-controlled friction law is appropriate and provides adequate results for the fingerpad-touchscreen contact (Heß and Forsbach, 2020).

Therefore, we use the classical Amontons-Coulomb friction law with a coefficient of friction (COF) for the fingerpad touchscreen contact of $\mu_0 = 0.25$ (Ayyildiz et al., 2018). In terms of the contact pressure $p$, this means a node sticks to the touchscreen if

$$\tau < \mu_0 p \tag{6}$$

and slips otherwise with

$$\tau = \mu_0 p. \tag{7}$$

It should be noted that the contact pressure is the sum of the elastic stress at the surface and the electrostatic attraction. Thus, electroadhesion directly increases the friction. Figure 2 shows the stress components for an exemplary case with electroadhesion for the frictional normal contact, the partially slipping tangential contact and the fully slipping tangential contact. Due to the large deformations of the fingerpad, there are significant point-symmetrical shear stresses for the normal contact and each ridge has small slipping areas at the contact edges. For the tangentially loaded fingerpad, slip propagates mostly from the leading edge of the contact.

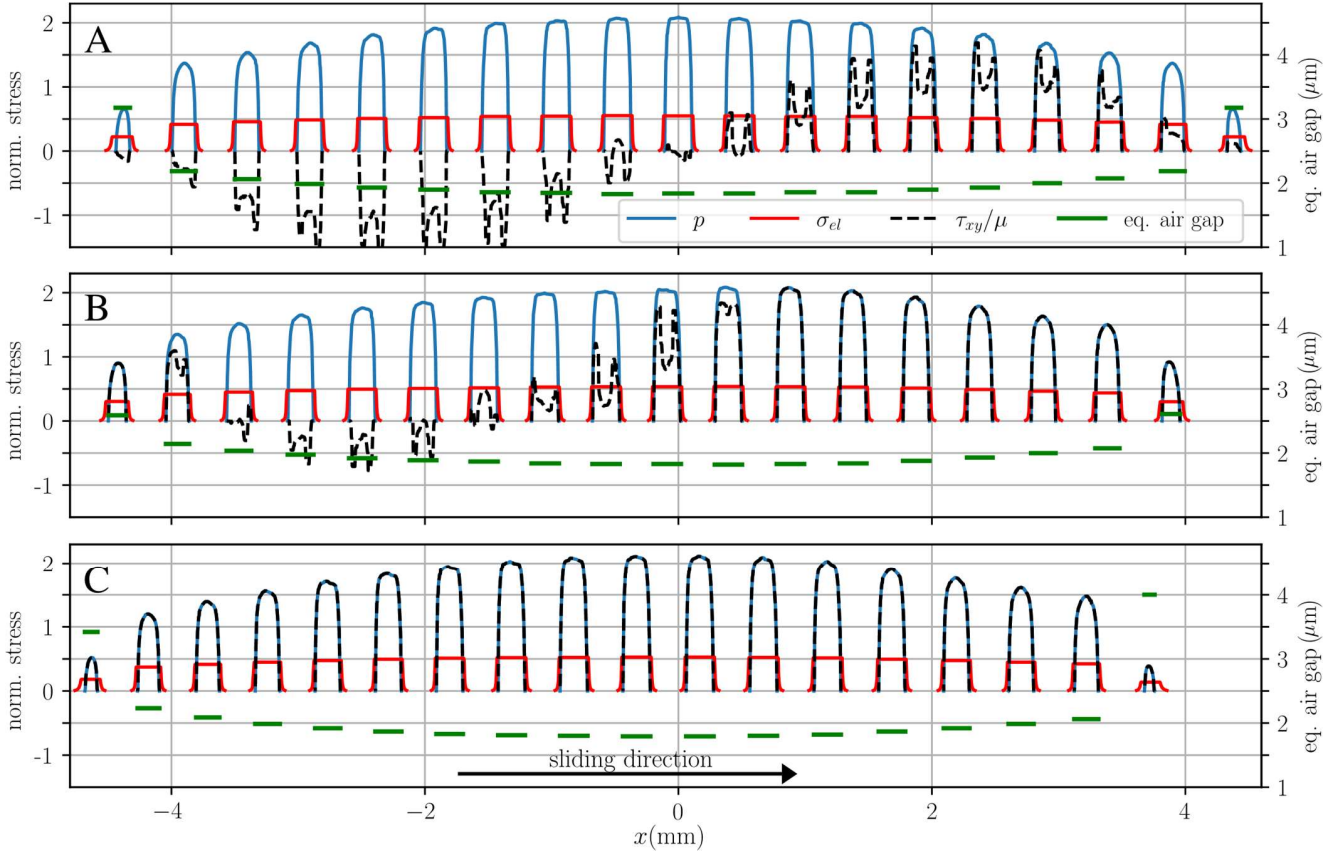

**Figure 2: Stress distributions in contact normalized with the average pressure in contact $f_{ext}/L_R$ and corresponding equivalent air gap at external force $f_{ext} = 0.12$N/mm and applied voltage $U = 150$V for the frictional normal contact (A), the partially slipping tangential contact (B) and the fully slipping tangential contact (C).**

## 3    Microscale roughness model

### 3.1    Fingerpad topography measurement and analysis

We conducted measurements on the fingerpad topography of a 25-year-old female using a rubber replica. The rubber replica was generated using a two-part silicone rubber compound (RepliSet-F1, Struers) that is directly applied to the fingerpad with a working life of less than 1 Minute. The topography of the replica was then measured using a 3D laser scanning microscope (Keyence VK-X100K). As the replica is a negative mold, the 3D height measurements were inverted and any macroscopic slope or curvature was removed. Figure 3 A shows exemplary results for 20x magnification and 100x magnification. For the microscale roughness model, the roughness on top of the ridges is of interest. We selected 8 spots of 100µm x 100µm on the ridge in between the sweat glands and measured the roughness profile with 100x magnification.

The averaged height distribution shown in Figure 3 B can be approximated by a Gaussian distribution with a RMS roughness of 3.1µm. We also computed the power spectral density (PSD) shown in Figure 3 C using the methods described in (Jacobs et al., 2017). The topography of interest is well described by a self-affine rough surface with a power-law PSD

$$C_{2D}(q) = 4\pi \frac{H}{1+H} \left( \frac{h_{rms}}{q_0} \right)^2 \left( \frac{q}{q_0} \right)^{-2-2H},$$ (8)

with the Hurst exponent $H = 0.75$, and the smallest wavevector $q_0 = 2\pi/\lambda_0 \approx 6.3 \cdot 10^5 \, 1/\text{m}$.

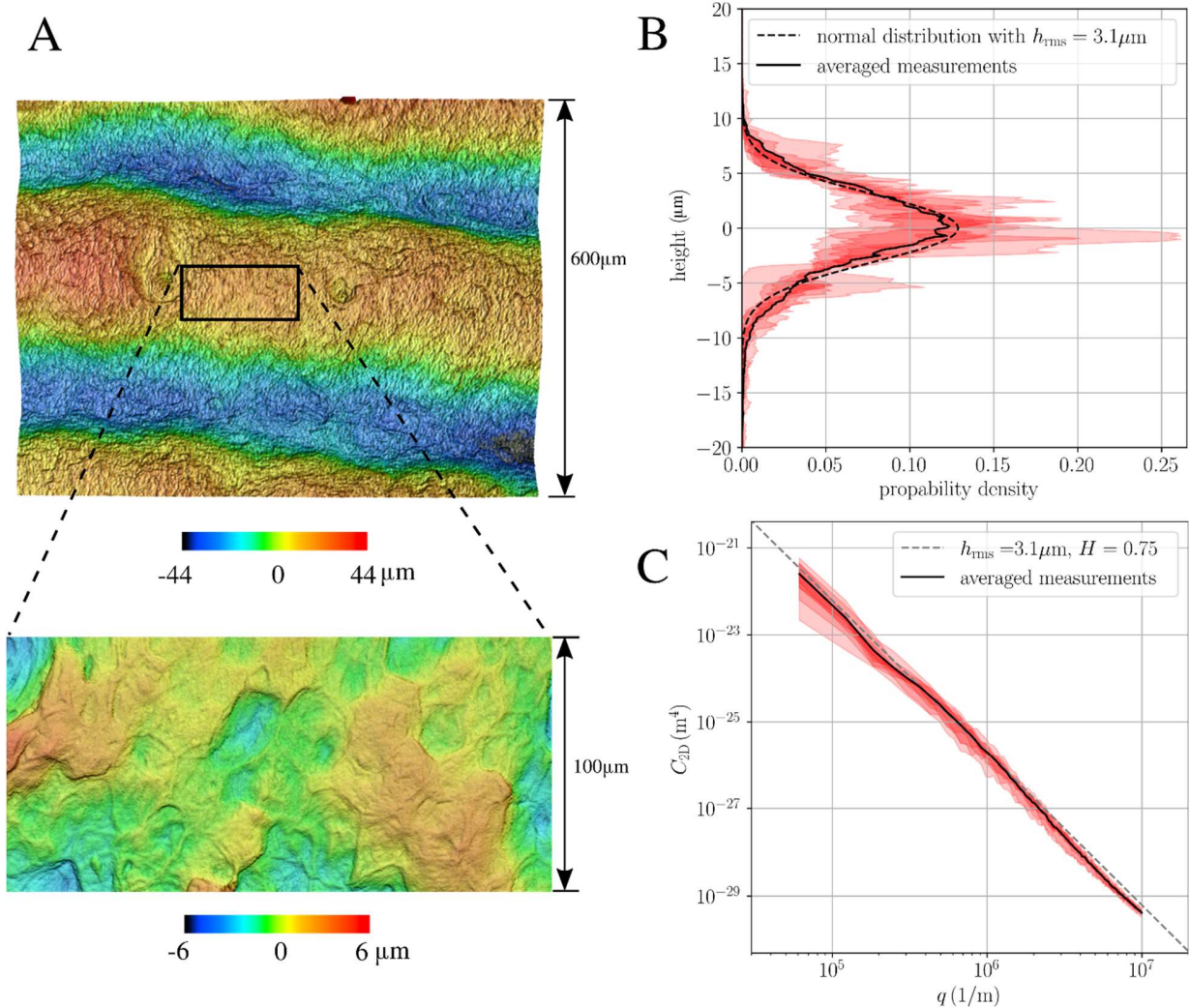

**Figure 3: Results of topography measurement. (A) Fingerpad with 20x magnification and 100x magnification. (B) Height distribution on top of the papillary ridges (100x magnification). (C) Two-dimensional PSD of 100µm x 100µm windows on top of the papillary ridges.**

## 3.2 Microscale model

We consider a rough ridge section of 100μm x 100μm in contact with a flat surface (the touchscreen). In agreement with the measurements in Section 3.1, we assume that

- the ridge exhibits a self-affine rough surface with a typical power-law PSD with Hurst exponent $H \approx 0.75$ (see Figure 3C),
- has a Gaussian height distribution (see Figure 3B)
- and its RMS Roughness is much smaller than the thickness of the outer skin layer (the stratum corneum layer, $h_{rms} \ll h_{SC}$).

In principle, any adhesive model on the microscopic scale can be used to determine the required relation of the external pressure on a papillary ridge and the equivalent air gap (see Eq. (5)). As a very simple estimate, we use the bearing area model (BAM) recently proposed by Ciavarella (2018) and formulated for electroadhesion by Ciavarella and Papangelo (2020). Given the assumptions above, the external pressure on the ridge in terms of the mean interfacial separation $\bar{u}$ may be approximated by

$$\bar{p}_{ext}(\bar{u}) \approx E_{SC}^* \, q_0 h_{rms} \exp\left(\frac{-\bar{u}}{\gamma h_{rms}}\right) - \frac{\varepsilon_0 U^2}{4 h_0^2}\left[\text{erfc}\left(\frac{\bar{u} - h_0}{\sqrt{2}\, h_{rms}}\right) - \text{erfc}\left(\frac{\bar{u}}{\sqrt{2}\, h_{rms}}\right)\right],$$ (9)

with the constant parameter $\gamma \simeq 0.5$ and the effective modulus of stratum corneum $E^*_{SC} = 4E_{SC}/3$ (incompressible), where $E_{SC} = 1\text{MPa}$ (Crichton et al., 2011). The first term in Eq. (9) is the repulsive pressure obtained by Persson's theory for intermediate mean separations (Persson, 2007). The second term is the adhesive contribution which is determined within the Maugis-Dugdale approximation by a constant electrostatic attraction

$$\sigma_0 = \frac{\varepsilon_0 U^2}{2 h_0^2}$$ (10)

in the adhesive area approximated by

$$\frac{A_{att}}{A_0} = \frac{1}{2}\left[\text{erfc}\left(\frac{\bar{u} - h_0}{\sqrt{2}\, h_{rms}}\right) - \text{erfc}\left(\frac{\bar{u}}{\sqrt{2}\, h_{rms}}\right)\right].$$ (11)

In Eq. (10), it was assumed that the range of the adhesive interaction is approximately the thickness of the effective insulating layer $h_0 = d_{sc}/\varepsilon_{r, sc} + d_i/\varepsilon_{r, i} \approx 0.5\mu m$ (see Table 3). We can obtain a relation between the equivalent airgap and the mean interfacial separation by equating the electrostatic attraction within the ridge contacts of the macroscopic model (Eqs. (3) and (5)) with the electrostatic attraction in Eq. (9):

$$\frac{\varepsilon_0 U^2}{2}\left(h_0 + d_{a, eq}\right)^{-2} = \frac{\varepsilon_0 U^2}{4 h_0^2}\left[\text{erfc}\left(\frac{\bar{u} - h_0}{\sqrt{2}\, h_{rms}}\right) - \text{erfc}\left(\frac{\bar{u}}{\sqrt{2}\, h_{rms}}\right)\right],$$ (12)

which simplifies to the geometric relation

$$d_{a, eq}(\bar{u}) = h_0\left\{\sqrt{2}\left[\text{erfc}\left(\frac{\bar{u} - h_0}{\sqrt{2}\, h_{rms}}\right) - \text{erfc}\left(\frac{\bar{u}}{\sqrt{2}\, h_{rms}}\right)\right]^{-1/2} - 1\right\}.$$ (13)

In Figure 4, this relation is shown for the parameters in Table 3. The equivalent air gap approaches a hard limit for small mean separations:

$$\min(\mathrm{d}_{a,eq}) = h_0 \left\{ \sqrt{2/\mathrm{erf}\left(\frac{h_0}{\sqrt{2}\,h_{rms}}\right)} - 1 \right\} \approx 1.49\,\mu m. \tag{14}$$

Finally, with Eqs. (9) and (13), the relation of external ridge loading to the equivalent air gap required for the macroscopic model (see Section 2.2) is derived. It is shown in Figure 5 for different applied voltages. Macroscopic adhesion only occurs for high voltages which is also reflected in the stickiness criterium obtained by Ciavarella and Papangelo (2020),

$$U > 1.8\,h_{rms}\sqrt{\frac{h_0 E^*}{\varepsilon_0 \lambda_0}} \approx 155\mathrm{V}\;. \tag{15}$$

Exemplary equivalent airgaps resulting from the coupling of the above relations of the microscopic to the macroscopic model are shown in Figure 2. The resulting equivalent airgaps for the considered loadings in the following section are mostly in the range $1.7\mu m - 3\mu m$ with slightly higher values at the outer ridges in contact. Similar values ($1.5\mu m$-$2.5\mu m$) are reported in a recent experimental study (Guo et al., 2019).

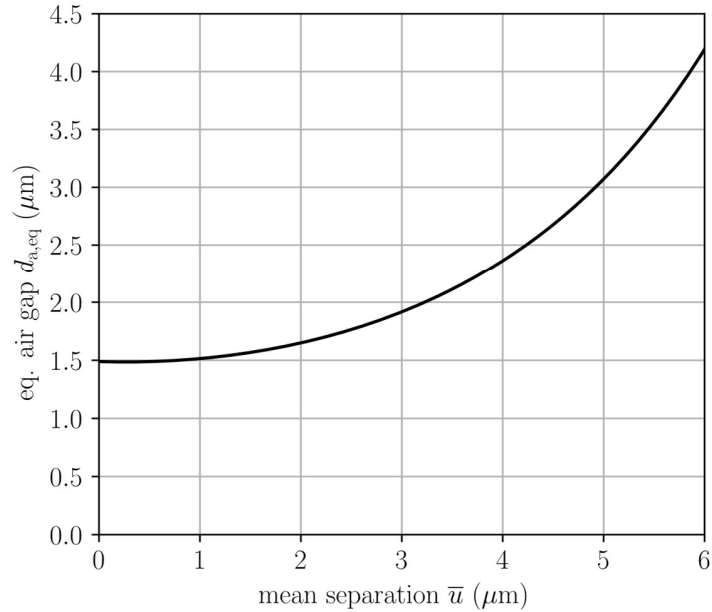

**Figure 4: Equivalent airgap $d_{a,eq}$ in terms of the mean interfacial separation $\overline{u}$.**

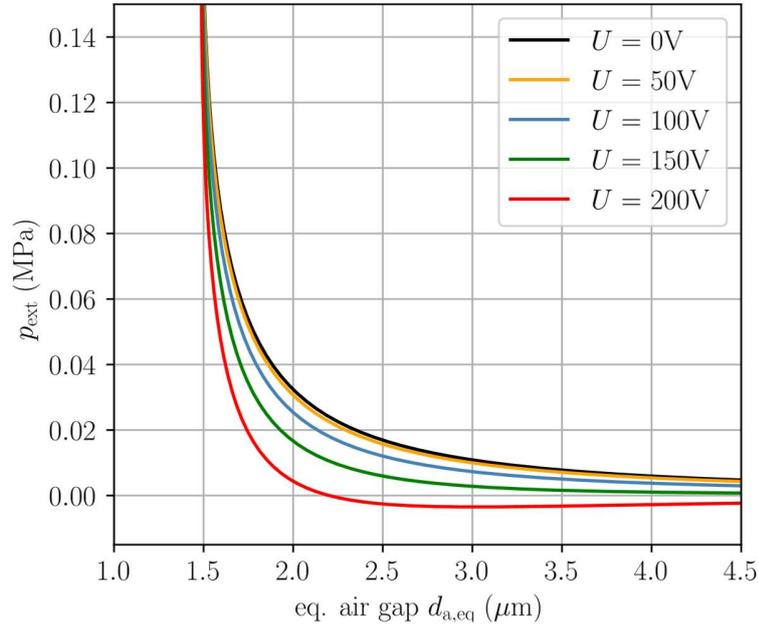

**Figure 5: External pressure $p_{\text{ext}}$ in terms of the equivalent airgap $d_{a,\text{eq}}$.**

**Table 3: Parameters used for the electroadhesive interaction**

| Symbol | Parameter Name | Value and Unit |
|--------|----------------|----------------|
| $\varepsilon_{r,sc}$ | Relative permittivity of stratum corneum | 1650 |
| $\varepsilon_{r,i}$ | Relative permittivity of insulating layer | 3.9 |
| $\varepsilon_{r,a}$ | Relative permittivity of air | 1 |
| $\varepsilon_0$ | Permittivity of free space | $8.854 \cdot 10^{-12}$As/Vm |
| $d_{sc}$ | Thickness of stratum corneum | 425μm |
| $d_i$ | Thickness of insulating layer | 1μm |
| $U$ | Applied voltage (peak to peak) | 0-200V |
| $h_{\text{rms}}$ | rms amplitude of roughness (microscale only) | 3.1μm |
| $E_{\text{sc}}$ | Elastic modulus of SC (microscale only) | 1 MPa |
| $\lambda_0$ | Largest wavelength (microscale only) | 100μm |

## 4     Results: The electroadhesive contact of a tangentially loaded fingerpad

Figure 6A shows the tangential force for the sliding fingerpad in terms of the applied external force for different applied voltages. For comparison with the experimental data by Ayyildiz et al. (2018), we assumed a fixed contact dimension of 20mm in out-of-plane direction for the two-dimensional model. The electrostatic attraction increases the contact pressure significantly (see also Figure 2C) and, thus, also the frictional force is increased as discussed in Section 2.3. <mark>A useful measure for electroadhesion is the ratio between the frictional force and the externally applied normal force, the "apparent" coefficient of friction (COF),</mark>

$$\mu_{\mathrm{app}} = \frac{F_{\mathrm{T}}}{F_{\mathrm{ext}}}, \tag{16}$$

<mark>shown in Figure 6B. For the electroadhesive frictional contact, this measure is generally higher than the actual COF in the Amontons-Coulomb friction law (see Eq. (7)) because the electrostatic attraction increases the normal contact pressure at the interface.</mark> As usual for adhesive contacts, the increase in apparent COF is particularly pronounced for small external forces. High voltages increase the tangential force and the apparent COF by more the 100%. For small voltages below 50V the electroadhesive effect will be negligible.

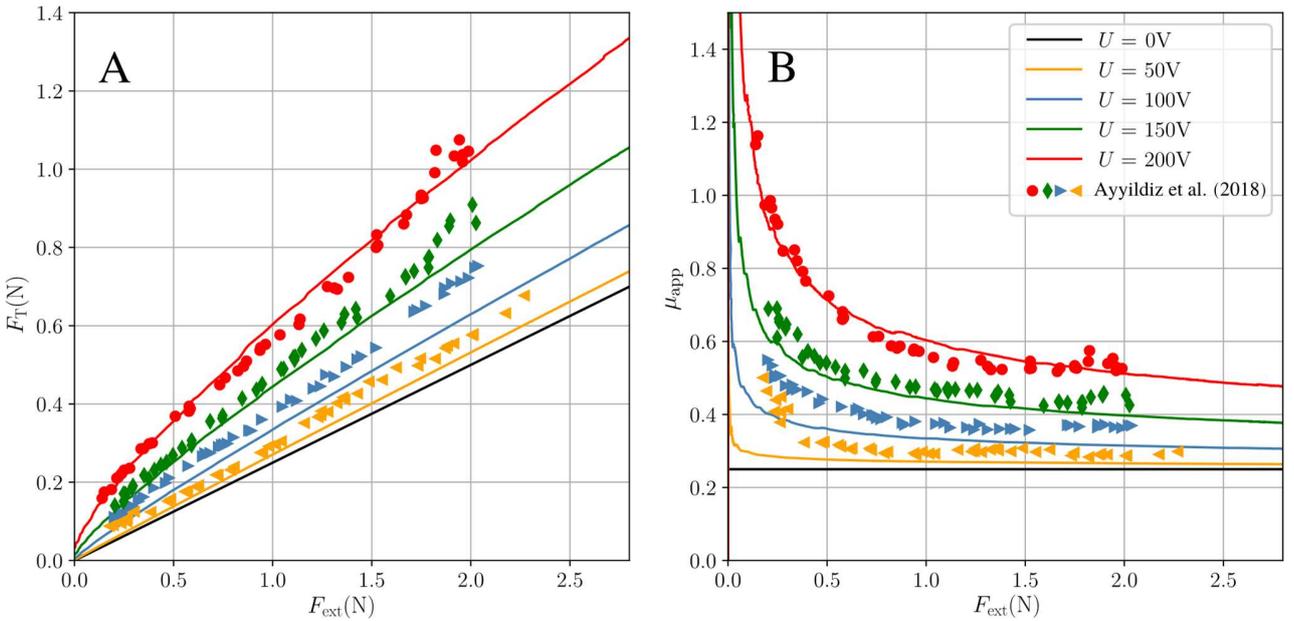

**Figure 6: Sliding fingerpad for different strengths of electroadhesion under the assumption of a fixed contact dimension of 20mm in out-of-plane direction. Markers show the experimental results by** Ayyildiz et al. (2018). **(A) Tangential force $F_{\mathrm{T}}$ in terms of external forces per unit length $F_{\mathrm{ext}}$; (B) Apparent coefficient of friction $\mu_{\mathrm{app}}$ in terms of external force $F_{\mathrm{ext}}$ .**

The trends concerning the tangential force and the apparent COF agree qualitatively and, for the out-of-plane dimension of 20mm, also quantitatively well with observations of recent experimental studies (Ayyildiz et al., 2018; Sirin et al., 2019). Of course, small discrepancies occur due to the simplifications of the complex three-dimensional problem. Most importantly, the plane strain assumption of the FE-model is a strong assumption. It is only accurate for a fingerpad in flat contact with the touchscreen but cannot account for the influence of the angle of finger and touchscreen that is usually used in experimental studies.

The influence of electroadhesion on the contact length is much smaller, but not negligible. Figure 7A shows the accumulated ridge contact length for the frictional normal and the sliding contact in case of the non-adhesive contact and the electroadhesive contact at 200V. Depending on the external force, the ridge contact length is increased by more than 20% for 200V. Note that this increase results from new ridges coming into contact as shown in Figure 7B and from the increase of the average contact length of the contacting ridges as shown in Figure 7C. The average ridge contact length is almost not affected by the contact state (stationary normal or sliding), but the number of contacting ridges is reduced by one or two for the sliding contact.

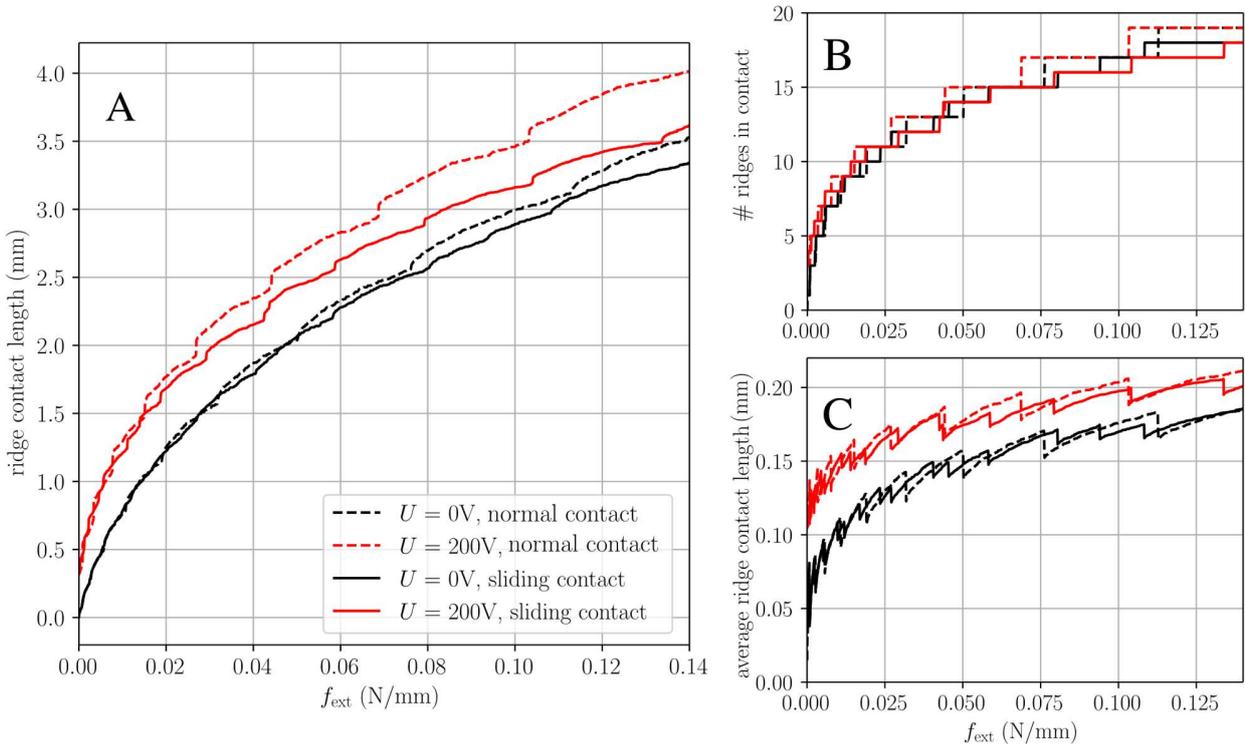

**Figure 7: (A) Accumulated ridge contact length for normal and sliding contact in terms of external forces $f_{ext}$ with and without electroadhesion; (B) Number of ridges in contact; (C) Average ridge contact length.**

The transition from normal contact to sliding contact for different applied voltages is shown in more detail in Figure 8A for a constant external force of 0.12N/mm. With electroadhesion, the contact length and the maximum applicable tangential force due to the Amontons-Coulomb friction law (see Eq. (6)) are increased. For increased tangential loading, the contact length is reducing in the global trend, but there is a significant waviness in the curves which can be easily explained by the large deformations of the fingerpad: Similar to a rolling motion (or rather a torsional deformation of the tissue around the bone whose rotation has been locked a priori), previously contacting ridges leave the contact at the trailing edge while new ridges come into contact at the leading edge (see Figure 2 and the contour plots in Figure 9A). Note that even without electroadhesion, the transition from pure normal contact to the onset of full slip is associated with a considerable reduction of the contact area. This effect is consistent with recent experimental and numerical studies on shear-induced contact area reduction of soft elastic materials (Sahli et al. (2018); Mergel et al. (2021)). The mechanism responsible for the area reduction under tangential loading has been described in previous studies (Heß and Forsbach, 2020; Lengiewicz et al., 2020); large deformations of the nonlinear elastic material induce a strong coupling of normal and tangential effects resulting in substantial vertical displacements (local lifting) and strain stiffening in the tangential direction. Adhesive interactions enhance the effect of area reduction, which is particularly large in percentage terms for small externally applied normal forces (Mergel et al., 2021). It should be noted that in the case of circular or elliptical contacts, both the experimentally observable reduction of the contact area and its anisotropic change can also be described by means of linear elastic fracture mechanics (Papangelo et al., 2019). The magnification of area reduction by turning on electrovibration predicted by our FE model (Figure 8A) is in good agreement with the experimentally recorded data of Sirin et al. (2019). However, due to the plane strain approach, our model cannot reproduce the anisotropic shear-induced change of the contact area, which was also observed in the aforementioned work under both conditions with and without electroadhesion.

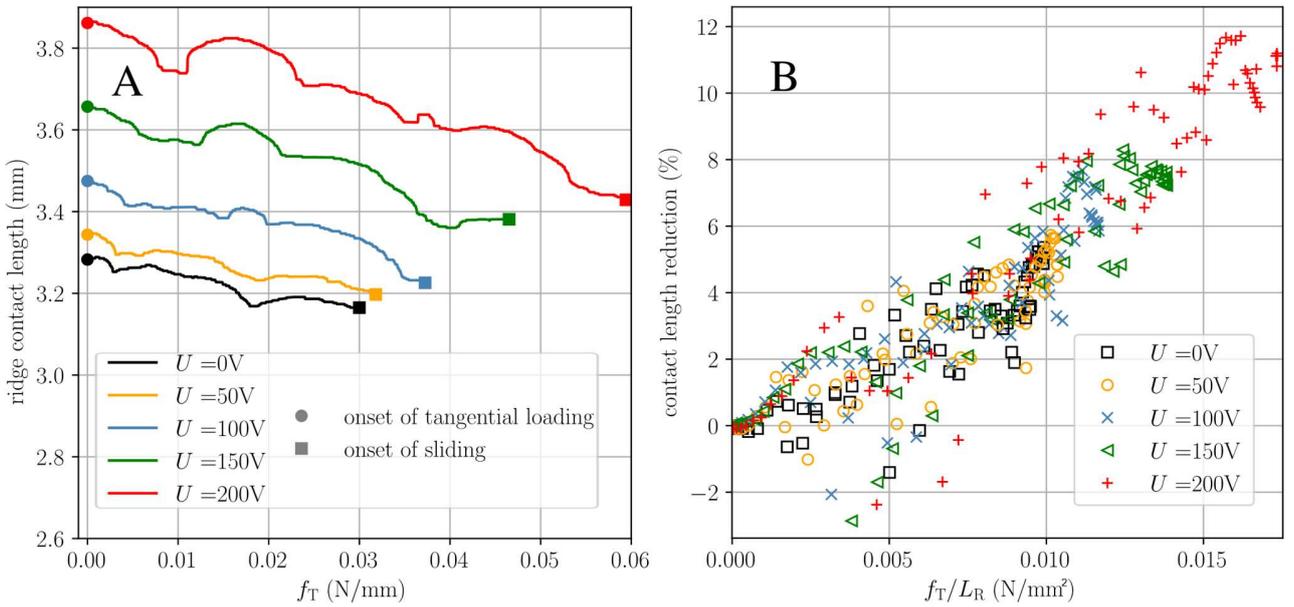

**Figure 8: (A) Ridge contact length in terms of the tangential force $f_T$ during the transition from stick to slip for $f_{\text{ext}} = 0.12\,\text{N/mm}$; (B) Reduction of ridge contact length in terms of the mean shear stress $f_T/L_R$.**

Figure 8B shows the contact length reduction, i.e., the reduction for the tangentially loaded contact compared to the normal contact at the same external force, in terms of the mean shear stress. The reduction, which is up to 12% for the simulated cases, depends approximately linearly on the mean shear stress and thus only indirectly on other parameters such as the applied voltage or the external force. The scattering of the data points again results from the discrete number of ridge contacts.

## 5 Discussion on tactile perception of electroadhesion

The tactile perception of the investigated fingerpad-touchscreen contact is conveyed by mechanoreceptive afferents in the skin. In response to spatiotemporally distributed mechanical quantities, the mechanoreceptors send pulses along the afferents to the brain. A common measure that correlates well with neural recordings in experiments is the *strain energy density* (SED) (Gerling and Thomas, 2008). The mechanism translating the mechanical stimulus to the firing of pulses is not yet fully understood and beyond the scope of the current work. However, depending on the type of mechanoreceptor, existing models use the magnitude of a mechanical stimulus, the rate of change, or a weighted mixture of both (Gerling et al., 2014). For now, we focus on the SED response at the sites of the Merkel receptors which are located in the tips of the intermediate ridges at the centers of the papillary ridges (see Figure 1). The numerous Merkel receptors belong to the slowly adapting mechanoreceptors (SA-I) with very localized receptive fields. Thus, they are partly responsible for the perception of friction and spatial changes in contact such as textures and roughness (Gardner and Martin, 2000). The response of mechanical stimuli like the SED under electroadhesion is thus very interesting for different applications including electroadhesion such as virtual textures.

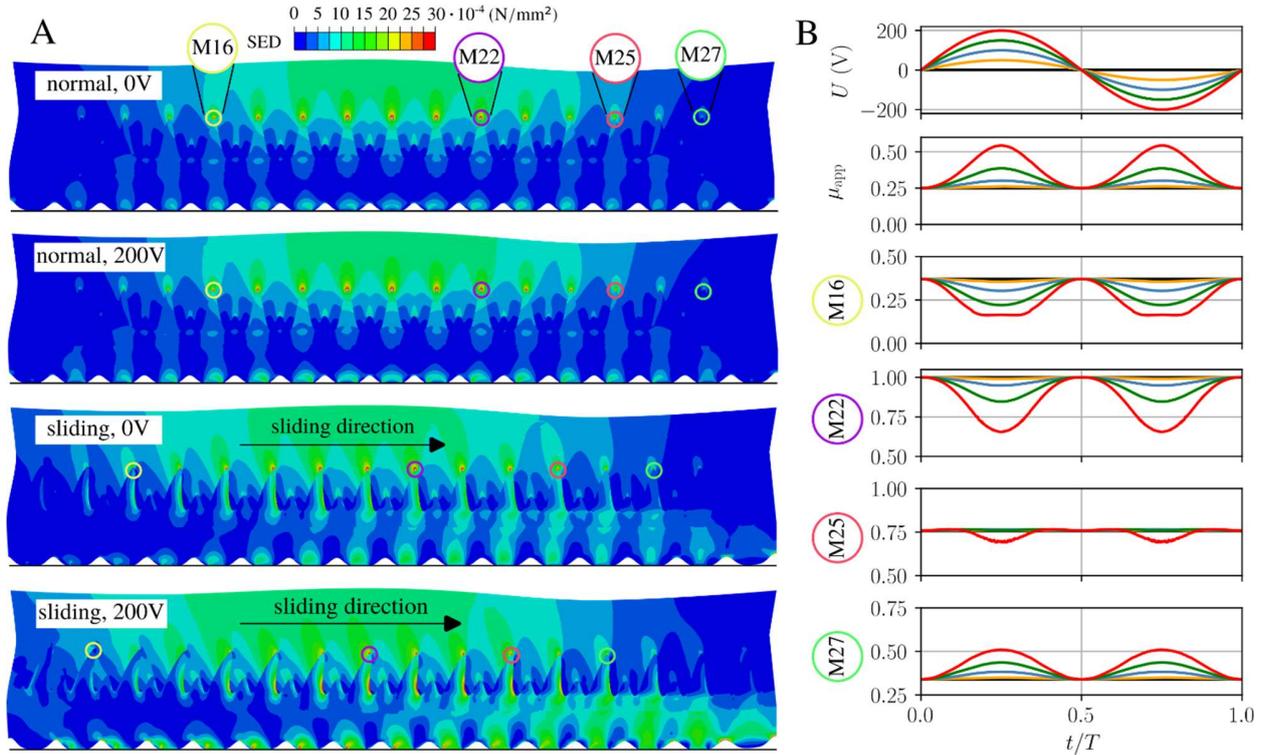

**Figure 9: (A) Contour plot of the strain energy density field at the mechanoreceptor sites for normal and sliding contact at $f_{ext} = 0.075\text{N/mm}$. (B) Apparent COF and normalized SED at different marked Merkel receptor sites in response to sinusoidal input voltage.**

In the context of the Ogden hyperelastic model, the SED is calculated as in Eq. (1). Figure 9A shows the distribution of the SED in the skin layers in vicinity of the contact for normal and sliding contact with and without electroadhesion. The epidermal-dermal junction is easily identified by the discontinuous SED due to the stiffness drop (see also Table 2). For the normal contact, the changes due to electroadhesion are almost completely limited to the stratum corneum layer; the local ridge contact area is increased as shown in Figure 7B and so is the local SED response. Since the receptor fields are located deeper in the skin, electroadhesion is not perceived in the normal contact which is also found in experiments (Ayyildiz et al., 2018). Maxima in the SED field are found at the Merkel receptor sites below the contacting ridges. For the sliding contact, there is a considerable tangential displacement which shifts the papillary ridges in contact. The intermediate ridges that are bedded in the soft surrounding dermis are bending which reduces the maximum SED values at most Merkel receptors and results in a second local SED maximum at the base of the intermediate ridges. If the electroadhesion is turned on during sliding, the trends described above are amplified. Considering that the discussed tactile perception of electroadhesion is only due to the increased tangential loading in the sliding fingerpad (we will exclude any vibratory perception through other receptor types for now), we can safely assume that the SED field and thus the tactile perception corresponds to the non-adhesive sliding fingerpad with increased COF of $\mu_0 = \mu_{app}$ ($\mu_0 \approx 0.55$ for the example in Figure 9A, see also Figure 6B).

It is interesting to note that the SED in the dermal papillae (see Figure 1) is increased for the tangentially loaded contact, because they are squeezed in between the bending intermediate ridges and the limiting ridges. The dermal papillae are the sites of the Meissner corpuscles; rapidly adapting mechanoreceptors (RA-I) that can detect spatially distributed changes in a mechanical stimulus.

Figure 9B shows the response of the apparent COF and the SED at single Merkel receptors to an applied sinusoidal input voltage. Note that we used a quasistatic model and thus, do not account for any explicit time dependencies such as viscous effects, wave propagation or leakage through the layers at low frequencies of the applied voltage. The SED values are normalized with the highest SED value in the receptor field for 0V (here at the receptor marked M22). Due the $U^2$ term in the electrostatic attraction (Eq. (3)), the apparent COF and the SED curves have twice the frequency of the input voltage. The transfer of apparent COF to the SED response is nonlinear and highly dependent on the receptor position: For some receptors the SED is increased or decreased because they are either entering the contact zone (M27) or leaving the contact zone (M16) and in some cases the SED is decreased due to the above-mentioned bending of the intermediate ridge (M22). A combination of the mechanisms may also result in almost no change in SED at all (M25). However, it should be noted that the tactile perception of vibrations strongly depends on their frequency spectrum. The Merkel receptors mediate the perception only for very low frequencies (<1.5Hz), while higher frequencies are mediated by Meissner (1.5-50Hz) or Pacinian corpuscles (>50Hz) (Gescheider et al., 2002).

## 6    Conclusion

We have developed a two-scale model for fingerpad friction under electroadhesion, employing a two-dimensional finite element model for the macroscopic scale and a simple bearing area model that accounts for the measured roughness on the papillary ridges. Both scales are coupled using the concept of an equivalent air gap. Unlike many models in the literature, the model avoids the use of matching parameters or unphysical simplifications; all quantities are taken from recent studies or, in case of the topography of the fingerpad, from our own measurements. The apparent coefficient of friction and the contact length (reduction) of the sliding electroadhesive fingerpad are in good qualitative agreement with recent experimental studies. Macroscopic adhesion is predicted only for very high voltages.

By using sufficiently detailed geometry and appropriate material models, the model will help to understand the underlying mechanisms of electroadhesion and, most importantly, can be readily used to simulate the effect of electroadhesion on tactile perception. We evaluated the strain energy density, a measure that correlates well with neural recordings, at different SA-I mechanoreceptor sites. For the stationary normal contact, the changes in the mechanical stimuli due to electroadhesion are small and not within the receptor fields. However, as expected, the increased friction in the sliding contact has a significant effect on the strain energy density response at the receptor fields. For a thorough investigation of the effect of electroadhesion magnitude and frequency on tactile perception, other receptor types (RA-I, RA-II) and a neural dynamics model need to be considered. This will be part of a future study addressing the tactile perception of virtual textures in comparison with their real counterparts.

Finally, the proposed FE-formulation is also of interest for simulating electroadhesion in soft robotics where there is often a need to control contact forces to perform gripping tasks and enhance holding capabilities (Mazzolai et al., 2019; Giordano et al., 2021a, 2021b). Just as in surface haptics, there is still a need for reliable models that can capture the complex electroadhesive interactions (Guo et al., 2019a).

# 8    Conflict of Interest

*The authors declare that the research was conducted in the absence of any commercial or financial relationships that could be construed as a potential conflict of interest.*

# 9    Author Contributions

Conceptualization, F.F. and M.H.; methodology, F.F., M.H. and A.P.; software, F.F.; validation, F.F.; formal analysis, F.F.; visualization, F.F.; writing—original draft preparation, F.F., M.H. and A.P.; writing—review and editing, F.F. and M.H.. All authors have read and agreed to the published version of the manuscript.

# 10    Funding

We acknowledge support by the German Research Foundation and the Open Access Publication Fund of TU Berlin.

AP was supported by the European Research Council through the ERC Starting Grant "Towards Future Interfaces With Tuneable Adhesion By Dynamic Excitation"— SURFACE, under the funding programme Horizon Europe (FP/2021–2027), ERC Grant Agreement ID: 101039198, DOI:10.3030/101039198. A.P. was supported by Regione Puglia (Italy), project ENOVIM (CUP: D95F21000910002) granted within the call "Progetti di ricerca scientifica innovativi di elevato standard internazionale" (art. 22 della legge regionale 30 novembre 2019, n. 52 approvata con A.D. n. 89 of 10-02-2021, BURP n. 25 del 18-02-2021). AP acknowledge support from the Italian Ministry of Education, University and Research (MIUR) under the program "Departments of Excellence" (L.232/2016).

## 11 Acknowledgments

We would like to thank Philip Köch for his cooperation and valuable discussions in the context of his bachelor thesis from 2021, where he performed some preliminary FE analyses.